\documentclass[twocolumn,showpacs,preprintnumbers,amsmath,amssymb]{revtex4-1}

\usepackage{graphicx}
\usepackage{dcolumn}
\usepackage{bm}
\usepackage{epsfig}
\begin{document}

\title{Observation of polaritons in Bi$_2$Sr$_2$CaCu$_2$O$_{8+x}$ single crystals}

\author{ Sven-Olof Katterwe}
\author{Holger Motzkau}
\author{Andreas Rydh}
\author{Vladimir M. Krasnov}

\email{Vladimir.Krasnov@fysik.su.se}

\affiliation{Department of Physics, Stockholm University, AlbaNova
University Center, SE -- 106 91 Stockholm, Sweden }




\begin{abstract}
The Bi$_2$Sr$_2$CaCu$_2$O$_{8+x}$ high-temperature superconductor
represents a natural metamaterial, composed of metallic CuO
bilayers sandwiched between ionic BiO planes. Each pair of CuO
bilayers forms an atomic-scale Josephson junction. Here
we employ the intrinsic Josephson effect for {\emph{in situ}}
generation and detection of electromagnetic waves in
Bi$_2$Sr$_2$CaCu$_2$O$_{8+x}$ single crystals. We observe that
electromagnetic waves form polaritons with several transverse
optical phonons. This indicates the presence of unscreened polar
response in cuprates, which may lead to strong electron-phonon
interaction. Our technique can provide intense local sources of
coherent, monochromatic phonon-polaritons with kW/cm$^2$ power densities.

\end{abstract}

\maketitle

Propagation of light in a media is accompanied by electromagnetic
(EM) interaction with charged particles. The interaction is strong
in polar materials, such as ionic insulators and ferroelectrics.
Resonances between light and EM-active collective modes in the
media (e.g., optical phonons \cite{Polaritons}, or excitons
\cite{Excitons1, Excitons2}) lead to formation of polaritons, half
light - half matter particles, with a singular dielectric function
\begin{equation}\label{Polariton}
    \epsilon_{\mathrm r}(\omega) = \epsilon_{\infty} + \sum_j \frac{\omega_{\mathrm{TO},j}^2 S_j}{\omega_{\mathrm{TO},j}^2-\omega^2 - i \gamma_j \omega}.
\end{equation}
It has poles at transverse $\omega_{\mathrm{TO}}$ and zeroes at
longitudinal $\omega_{\mathrm{LO}}$ optical frequencies. Here
$S_j$ and $\gamma_j$ are oscillator strengths and damping
parameters, respectively. Light does not propagate through the
media at $\omega_{\mathrm{TO}} < \omega < \omega_{\mathrm{LO}}$
for the same reason that low frequency EM waves do not propagate
through metals, namely due to the negative dielectric constant
Re$(\epsilon_{\mathrm r})<0$. For $\gamma \rightarrow 0$ the group
velocity of light turns to zero at $\omega_{\mathrm{TO}}$. The
ability of polaritons to slow down, or even stop and store photons
has attracted significant attention in recent years because it may
be used in optical metamaterials \cite{Polaritonics,Zheludev} and
facilitate manipulation of photons in optoelectronics
\cite{Zayats}.

Here we study EM-wave propagation in the layered high temperature
superconductor Bi$_2$Sr$_2$CaCu$_2$O$_{8+x}$ (Bi-2212).
Superconductivity in cuprates occurs as a result of doping of the
parent ionic oxide insulator. Optical spectroscopy has revealed an
intermixture of metallic (Drude-type anomaly at $\omega = 0$) and
ionic [strong infrared (IR) phonons] properties in cuprates
\cite{Basov,VanDerMarel,Kovaleva}. However, Bi-2212 is not a
uniform doped Mott-insulator. It represents a natural atomic
superlattice, composed of metallic CuO$_2$-Ca-CuO$_2$ bilayers
sandwiched between ionic SrO-2BiO-SrO planes. Most clearly this
was demonstrated by observation of the intrinsic Josephson effect
between CuO bilayers \cite{Kleiner94,Munzar,Katterwe}. From an
optical point of view, Bi-2212 represents a layered photonic
crystal \cite{Ooi1, Wu2, Savelev3} -- a multilayer transmission
line for EM waves. Therefore, low frequency transverse EM (TEM)
waves can propagate along insulating BiO planes
\cite{Superluminal} despite that Bi-2212 is metallic and
superconducting. Importantly, the intrinsic Josephson effect
allows both {\it in situ} generation and detection of TEM waves in
the crystal, i.e., the material as such can convert the dc-bias
current into high frequency ac-electric field via the quantum
mechanical ac-Josephson effect.

We study small mesa structures, fabricated on top of Bi-2212
single crystals using micro/nano-fabrication techniques. We use
two batches of crystals: pure, slightly underdoped Bi-2212 with
critical temperature $T_{\mathrm c} \simeq 82$\,K, and lead-doped,
slightly overdoped Bi$_{2-y}$Pb$_y$Sr$_2$CaCu$_2$0$_{8+x}$ with
$T_{\mathrm c} \simeq 87$--$91$\,K. The most noticeable difference
between them is in the $c$-axis critical current density
$J_{\mathrm c} (4.2\,\mathrm{K})\simeq 10^3$ and $10^4$\,A/cm$^2$
for pure and lead-doped mesas, respectively. Measurements were
performed in a gas-flow $^4$He cryostat in a temperature range
down to 1.6\,K and magnetic field $H$ up to 17\,T. Samples were
mounted on a rotatable sample holder. Accurate alignment of the
field parallel to CuO planes was needed to avoid intrusion of
Abrikosov vortices in intense magnetic fields. For alignment we
used the procedure developed in Ref.~\cite{Katterwe}, which
facilitates alignment with an accuracy better than 0.02$^{\circ}$.
Details of sample fabrication and measurements can be found in
Refs.~\cite{Katterwe} and \cite{Superluminal}.

\begin{figure}
\begin{center}
\includegraphics[width=0.9\linewidth]{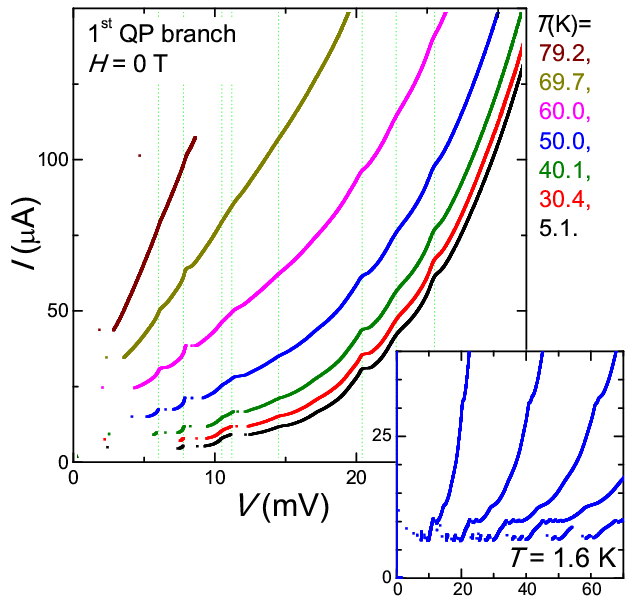}
\end{center}
\caption{Temperature dependence of the first QP
branch for the lead-doped mesa~M1 in zero magnetic field. It is seen that phonon
frequencies are $T$-independent. The inset shows QP branches in the
$I$-$V$ of the same mesa at $T = 1.6$\,K. Resonances with LO
phonons are seen at every QP branch.}
\label{Fig1}
\end{figure}

The inset of Fig.~\ref{Fig1} shows current-voltage ($I$-$V$)
characteristics of a lead-doped mesa (M1, $1.3 \times 1.9
\,\mu\mathrm{m}^2$). Periodic branches are
due to one-by-one switching of junctions from the superconducting
to the quasiparticle (QP) state \cite{Kleiner94,SecondOrder}.
Additional small sub-branches are seen at every QP branch. They
represent resonances between the ac-Josephson oscillations in
intrinsic junctions and $c$-axis LO phonons
\cite{Schlen,Ponomarev,Yurgens,Heim}. Figure~\ref{Fig1} shows
temperature dependence of the first QP branch for the same mesa.
Unlike the superconducting gap and QP resistance, which decrease
with increasing $T$ \cite{SecondOrder}, the voltages of phonon
resonances remain independent of $T$ and correspond to frequencies
of IR and Raman active $c$-axis phonons
\cite{VanDerMarel,Kovaleva,Raman}. Their
identification is provided in Table~\ref{Table1}.

The amplitude of phonon resonances is proportional to $J_{\mathrm
c}^2$ \cite{Heim}. Since $J_{\mathrm c}$ for our lead-doped
crystals is about ten times larger than for pure Bi-2212 crystals,
phonon resonances in the lead-doped mesas are greatly enhanced.
For analysis of resonances in pure Bi-2212 we have to look on
$\mathrm{d}I/\mathrm{d}V$ characteristics \cite{Yurgens}.
Similarly, resonances are strongly suppressed by a moderate
$c$-axis magnetic field. This is caused by penetration of
Abrikosov pancake vortices in CuO planes, which suppress
superconductivity \cite{KrMag} and $J_{\mathrm c}$. In contrast,
in-plane magnetic field up to 17\,T is not affecting
superconductivity and $J_{\mathrm c}$ at $T \ll T_{\mathrm c}$ due
to the extremely large anisotropy of Bi-2212. The apparent
Fraunhofer modulation of the $c$-axis critical current in this
case \cite{Katterwe} is caused by interference of opposite
supercurrents, circulating in Josephson fluxons, with unsuppressed
amplitude $J_{\mathrm c}$.

At sufficiently large magnetic field, fluxons form a regular
lattice in the mesa \cite{Katterwe}. The number of fluxons per
junction is given by the ratio $\Phi/\Phi_0$, where $\Phi=H L s$
is the flux per junction, $L$ is the junction length, $s\simeq
1.5$\,nm is the stacking periodicity of intrinsic junctions, and
$\Phi_0$ is the flux quantum. The $c$-axis bias current exerts a
Lorentz force and causes unidirectional fluxon motion. This leads
to appearance of an additional flux-flow branch in the $I$-$V$
curve, with the voltage per junction
$V_{\mathrm{FF}}=u_{\mathrm{FF}}Hs$, where $u_{\mathrm{FF}}$ is
the fluxon velocity. Periodic collision of fluxons with the edge
of the mesa leads to flux-flow oscillations \cite{Superluminal}.
Produced EM waves are partly emitted and partly reflected back
into the junctions. The latter start to travel as TEM waves in the
multilayered transmission line formed by the mesa, get reflected
at the opposite edge, and continue bouncing until decay.

\begin{table}%
\begin{tabular}{cccccc}%
       \hline%
       $\#$& $V$ & $\omega_{\mathrm{LO}}$  &  type& symmetry& assignment\\
       & (mV) & $({\mathrm{cm}^{-1}})$ & & & \\
       \hline
       1 & 6.0 & 97 & IR & A$_\mathrm{2u}$ & Bi':Cu1CaSr \\
       2 & 7.8 & 126 & Raman & A$_\mathrm{1g}$ & Cu1Sr \\
       3 & 10.5 & 169 & IR & A$_\mathrm{2u}$& Sr:Cu1' \\
       4 & 11.2 & 181 & Raman& A$_\mathrm{1g}$& Sr:Cu1' \\
       5 & 14.5 & 234 & IR & A$_\mathrm{2u}$& Ca:Sr' \\
       6 & 20.4 & 329 & IR & A$_\mathrm{2u}$& O3O1 \\
       7 & 22.8 & 368 & IR & A$_\mathrm{2u}$& O1':CaO3 \\
       8 & 25.5 & 411 & Raman & A$_\mathrm{1g}$& O1:Sr' \\
       \hline
       \end{tabular}
\caption{\label{Table1}%
\sffamily\footnotesize\noindent{Identification of LO phonon resonances at zero field, marked in Fig.1 (following Ref.~\cite{Kovaleva}).}}
\end{table}

\begin{figure}
\begin{center}
\includegraphics[width=0.8\linewidth]{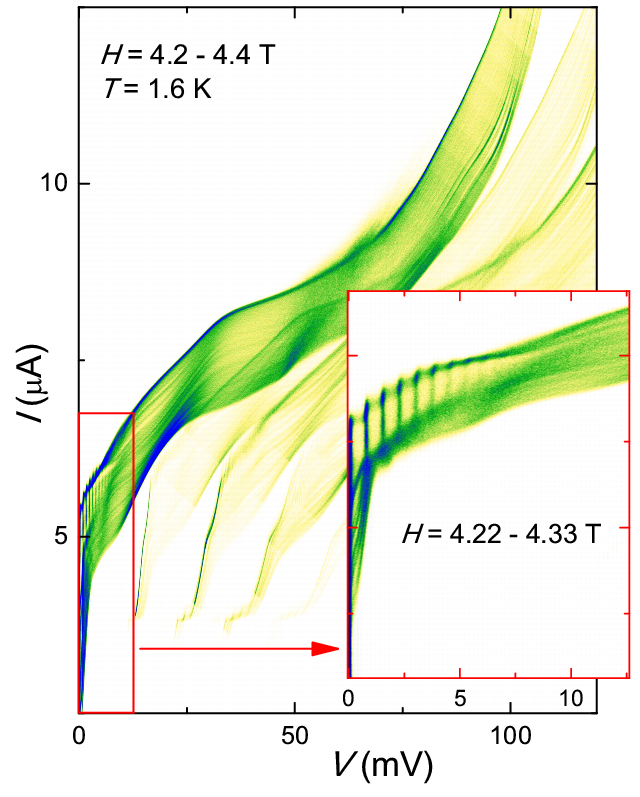}
\end{center}
\caption{Flux-flow part of the $I$-$V$ for the
lead-doped mesa~M2, measured during continuous sweep of the
in-plane magnetic field from 4.2 to 4.4\,T. The inset shows a
series of Fiske steps, caused by formation of standing waves
inside the mesa. It demonstrates the Josephson flux-flow mechanism
of TEM wave generation in our experiment.}
\label{Fig2}
\end{figure}

Figure~\ref{Fig2} shows flux-flow part of the $I$-$V$ for a
lead-doped mesa (M2, $1.2 \times 2\,\mu\mathrm{m}^2$), measured
during continuous sweep of the in-plane magnetic field from 4.2 to
4.4\,T. The flux-flow voltage grows gradually with $H$. Bouncing
TEM waves can be explicitly visualized at geometrical resonance
conditions, corresponding to formation of standing waves in the
mesa. They are seen as a series of Fiske steps on the flux-flow
branch \cite{Superluminal}. We note that those TEM waves are
photon (or surface plasmon \cite{Zayats}) like, because they have
a linear dispersion relation at low frequencies. This leads to
equidistance of Fiske steps \cite{Superluminal}. The inset in
Fig.~\ref{Fig2} demonstrates the Fiske step sequence,
corresponding to out-of-phase oscillations in neighboring
junctions with one node per junction \cite{Superluminal}. The
corresponding slowest speed of light $c_N \simeq 3.9\times
10^5$\,m/s is almost one thousand times slower than the speed of
light in vacuum -- not because of the extraordinary large
$\epsilon_{\mathrm r}$, but due to an extraordinary large kinetic
inductance of the atomically thin intrinsic junctions.

The strongest geometrical resonance occurs at the
velocity-matching condition, when $u_{\mathrm{FF}}$ is equal to
the phase velocity of TEM waves \cite{Superluminal}. In this case
waves are synchronized with the fluxon chain and the wave length
is equal to the separation between fluxons. The corresponding TEM
wave number is
\begin{equation}\label{kTEM}
k = 2 \pi H s /\Phi_0.
\end{equation}
It can be easily tuned by changing $H$. In experiment the velocity
matching resonance is seen as a sharp upturn (maximum in
d$I$/d$V$) at the top of the flux-flow branch. The corresponding
frequency is given by the ac-Josephson relation
\begin{equation}\label{OmegaFF}
\omega = 2\pi V_{\mathrm{FF}}^* / \Phi_0,
\end{equation}
where $V_{\mathrm{FF}}^*$ is the measured maximum flux-flow
voltage per junction. Eqs. (\ref{kTEM},\ref{OmegaFF}) provide the
basis for analysis of the dispersion relation in this work.

\begin{figure}
\begin{center}
\includegraphics[width=0.8\linewidth]{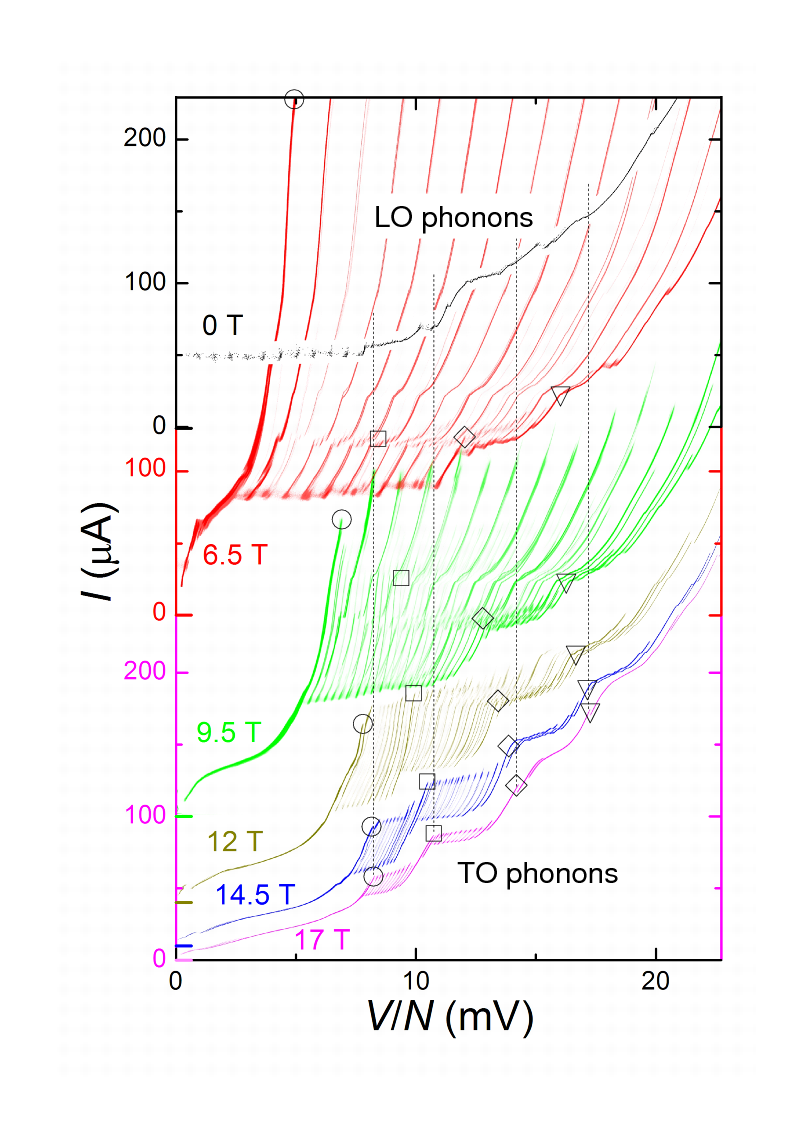}
\end{center}
\caption{A set of $I$-$V$ curves for the
lead-doped mesa~M3 at different in-plane magnetic fields and
$T=1.6$\,K. The curves are offset vertically for clarity.
Development of the velocity-matching flux-flow resonance (circles)
and higher flux-flow/phonon resonances (squares, rhombuses and
triangles) is seen. With increasing $H$ voltages of
flux-flow-phonon resonances first increase and then saturate when
the ac-Josephson frequency is approaching the next TO phonon
frequency (marked by vertical lines).} \label{Fig3}
\end{figure}

Figure~\ref{Fig3} shows $I$ vs. $V$ per junction at different
in-plane magnetic fields for a lead-doped mesa (M3, $0.9\times
1.3\, \mu\mathrm{m}^2$) with $N =11$ junctions. The
velocity-matching flux-flow resonance is marked by circles. It is
moving to higher voltages with increasing $H$ and eventually
saturates upon approaching the voltage of phonon $\#2$ in
Table~\ref{Table1}. Above it we observe a series of additional,
remarkably strong resonances. The first three are marked by
squares, rhombuses and triangles, respectively. They develop
gradually out of one of the LO phonon resonances at $H=0$, grow in
amplitude and move to higher voltages with increasing field.
Finally, they saturate at high fields when the voltage is
approaching the next phonon resonance, as indicated by vertical
lines. Those resonances can not be associated with bare LO
phonons, which are field independent, but represent a combination
of flux-flow and phonon resonances. Each series have $N$
sub-branches with voltages $V_j=(N-j)V_{\mathrm{FF}}+jV_{\mathrm
R}$, ($j=1,2,\ldots N$) due to one-by-one switching of $N$
junctions from the flux-flow $V_{\mathrm{FF}}$ to the resonance
state $V_{\mathrm R}$. $N=11$ sub-branches in each series can be
distinguished at $H=17$\,T in Fig.~\ref{Fig3}. The last sub-branch
in each series represents the flux-flow/phonon resonance
frequency.

\begin{figure}
\begin{center}
\includegraphics[width=0.95\linewidth]{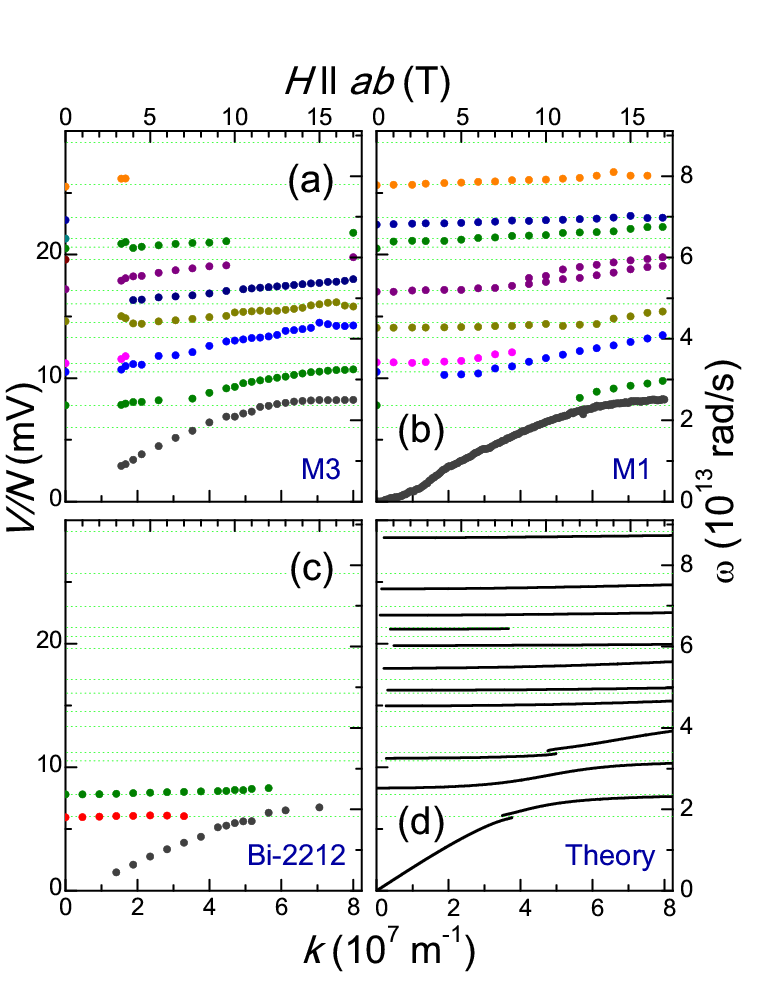}
\end{center}
\caption{Obtained TEM-wave dispersion relations
for (a) mesa~M3, (b) mesa~M1, and (c) a pure Bi-2212 mesa.
It is seen that the dispersion relation is linear at low
frequencies (constant speed of light), but saturate (zero group
velocity) whenever the frequency is approaching the phonon
frequency, indicating formation of polaritons. Panel~(d) shows
simulated polaritonic dispersion relation from Eq.~(\ref{Polariton}).
Horizontal lines represent all observed LO frequencies.}
\label{Fig4}
\end{figure}

Figure~\ref{Fig4} presents our main result: voltage per junction
of flux-flow/phonon resonances vs. $H$, which are translated into
the dispersion relation $\omega$ vs. $k$ via
Eqs.~(\ref{kTEM},\ref{OmegaFF}). Figure~\ref{Fig4}(a) shows the
dispersion relation for the mesa~M3. The lowest dispersion branch
corresponds to the regular flux-flow branch, indicated by circles
in Fig.~\ref{Fig3}. It is roughly linear at low $k$, corresponding
to a constant speed of light $\simeq 4\times 10^5$\,m/s close to
the out-of-phase TEM velocity, obtained independently from
analysis of Fiske steps \cite{Superluminal}. At larger $k$
($H>12$\,T) we observe a clear saturation with the group velocity
d$\omega$/d$k \rightarrow 0$ when the flux-flow frequency is
approaching the lowest Raman active phonon $\#2$ in
Table~\ref{Table1}. Higher dispersion branches correspond to
flux-flow/phonon resonances, indicated in Fig.~\ref{Fig3}. They
start flatly from a certain LO phonon frequency at $k=0$, grow at
intermediate $k$ and saturate again at large $k$, when the
frequency approaches the next phonon (indicated by horizontal
lines). Figure~\ref{Fig4}(b) demonstrates a similar behavior of
mesa~M1, which is slightly more overdoped and has about 30$\%$
larger $J_{\mathrm c}$. Figure~\ref{Fig4}(c) shows the dispersion
relation obtained for a pure Bi-2212 mesa. Since phonon resonances
are much weaker here due to smaller $J_{\mathrm c}$, we are able
to trace interaction only with the two lowest phonons.

The obtained results are in excellent agreement with theoretical
prediction by Preis {\it et\,al.} \cite{Preis} on the mutual
influence of flux-flow and phonon resonances in Josephson
junctions with ionic barriers. Interaction of oscillating electric
field with IR and Raman active phonons leads to formation of
polaritons, which slow down the group velocity of light to zero
$\mathrm{d}\omega /\mathrm{d}k \rightarrow 0$  at $\omega
\rightarrow \omega_{\mathrm {TO}}$ and prohibit propagation of TEM
waves and flux-flow at voltages $\Phi_0 \omega_{\mathrm{TO}}/2\pi
< V_{\mathrm{FF}} < \Phi_0 \omega_{\mathrm{LO}}/2\pi$. This leads
to the observed stratification of $I$-$V$ characteristics in
Fig.~\ref{Fig3}. Our data is well described by the polaritonic
dispersion relation of Eq.~(\ref{Polariton}), as shown in
Fig.~\ref{Fig4}(d).

Observation of strong polariton-like interaction of TEM waves with
IR and Raman active $c$-axis phonons highlight the unusual natural
metamaterial-type electronic structure of cuprates, in which
two-dimensional metallic CuO layers are sandwiched between ionic
layers. The combination of long range Coulomb interaction in
two-dimensional metallic layers with polar media around them, may
lead to an unusually strong electron-phonon interaction
\cite{Alexandrov5, Lanzara1, Falter2, Lee3, Bishop4}. This is
import for understanding of the puzzling phenomenon of high
temperature superconductivity.

The maximum power, pumped into phonon-polariton emission can be
explicitly estimated from the amplitude of flux-flow/phonon
resonances in Fig. 3: $P=\Delta I V_R \sim 10\,\mu$W, where
$\Delta I \sim 100\,\mu$A is the maximum resonance amplitude and
$V_R \sim 100$\,mV is the total voltage at the resonance. This
corresponds to kW/cm$^2$ emission power density. It can be further
enhanced in a coherent (cascade) manner using mesas with larger
amount of stacked junctions \cite{TheoryFiske}. Thus, our results
demonstrate that voltage biased Bi-2212 mesas act as compact and
intense source of coherent, monochromatic phonon-polaritons (or
surface plasmons) with power densities $\sim$kW/cm$^2$. Such local
surface-plasmon sources are on demand in a rapidly growing area of
optoelectronics \cite{Polaritonics,Zheludev,Zayats}. The
discovered phenomenon may also be used for creation of a THz
phonon laser \cite{TheoryFiske}.

{\bf Acknowledgments} The work was supported by the Swedish
Research Council, the K.~\&~A. Wallenberg foundation and the
SU-Core Facility in Nanotechnology.



\end{document}